\documentclass[journal=jacsat,manuscript=article]{achemso}

\usepackage[version=3]{mhchem} 
\usepackage{amsthm,amsfonts,amssymb,amscd,mathrsfs,xspace,framed}
\usepackage{color}

\author{Liang Zhang}
\email{zhangliang@semi.ac.cn}
\affiliation{Research and Development Center for Wide Bandgap Semiconductors, Institute of Semiconductors, Chinese Academy of Sciences, Beijing 100083, China}
\alsoaffiliation{College of Materials Science and Opto-Electronic Technology, University of Chinese Academy of Sciences, Beijing 100049, China}

\author{Yanan Guo}
\email{ynguo@semi.ac.cn}
\affiliation{Research and Development Center for Wide Bandgap Semiconductors, Institute of Semiconductors, Chinese Academy of Sciences, Beijing 100083, China}
\alsoaffiliation{College of Materials Science and Opto-Electronic Technology, University of Chinese Academy of Sciences, Beijing 100049, China}

\author{Junxi Wang}
\affiliation{Research and Development Center for Wide Bandgap Semiconductors, Institute of Semiconductors, Chinese Academy of Sciences, Beijing 100083, China}
\alsoaffiliation{College of Materials Science and Opto-Electronic Technology, University of Chinese Academy of Sciences, Beijing 100049, China}

\author{Jinmin Li}
\affiliation{Research and Development Center for Wide Bandgap Semiconductors, Institute of Semiconductors, Chinese Academy of Sciences, Beijing 100083, China}

\author{Jianchang Yan}
\email{yanjc@semi.ac.cn}
\affiliation{Research and Development Center for Wide Bandgap Semiconductors, Institute of Semiconductors, Chinese Academy of Sciences, Beijing 100083, China}
\alsoaffiliation{College of Materials Science and Opto-Electronic Technology, University of Chinese Academy of Sciences, Beijing 100049, China}
\title
  {3D Heterogeneous Integration of Silicon Nitride and Aluminum Nitride on Sapphire toward Ultra-wideband Photonics Integrated Circuits}

\begin{document}







\begin{abstract}
 Extending two-dimensional photonic integrated circuits (PICs) to three-dimensional (3D) configurations promises great potential for scaling up integration, enhancing functionality, and improving performance of PICs. Silicon-based 3D PICs have made substantial progress due to CMOS compatibility. However, the narrow bandgap of silicon (1.1 eV) limits their use in short-wavelength applications, such as chemical and biological sensing, underwater optical communications, and optical atomic clocks. In this work, we developed a 3D photonics platform by heterogeneously integrating silicon nitride (SiN) and aluminum nitride (AlN) PICs on sapphire (Al$_2$O$_3$). The broadband transparency of these materials allow our platform to operate over a multi-octave wavelength ranging from ultraviolet to infrared. Leveraging this platform, we demonstrated efficient optical nonlinearity in an AlN microcavity, low-loss and tunable SiN waveguide-based optical components, and optical linking between AlN and SiN PICs layers in the visible and near-infrared spectrum, hinting at potential applications in integrated quantum systems. Our work presents an ultra-wideband 3D PICs platform, providing new opportunities for broadband and short-wavelength applications of PICs.
\end{abstract}

\section{Introduction}
Photonics integrated circuits (PICs) have been widely developed and deployed, offering integrated solutions for various applications including telecommunications, computation, detection, and sensors~\cite{chovan2018photonic,shen2017deep,bogaerts2020programmable,najafi2015chip,goossens2017broadband}. Ever-increasing demands for data processing capabilities drive the need for large-scale PICs with more complex and numerous components~\cite{xu2024large,bao2023very,zhang2022large,bruckerhoff2023large}. For traditional two-dimensional (2D) planar PICs, integrating a large number of photonic devices on a limited chip area generally necessitates many optical waveguide crossings~\cite{xu2024large,bao2023very,zhang2022large}. This, in turn, results in serious in-plane crossing loss~\cite{liu2014ultra}, limiting the scalability of PICs. Inspired by 3D electronic integrated circuits (ICs) that leverage multiple metal interconnect levels for on-chip connectivity~\cite{reif2002fabrication,topol2006three}, extending 2D PICs to 3D structures through vertical stacking and coupling, provides a feasible solution for compact and highly dense large-scale PICs~\cite{biberman2011photonic,shang2015low,zhang2020scalable}.

Harnessing established CMOS manufacturing, silicon-based 3D PICs have achieved significant advancements, enabling demonstrations such as large-scale optical phased arrays~\cite{kim2019single}, efficient optical transceivers~\cite{chang20233d}, high-density optical switches~\cite{suzuki2020nonduplicate}, and integrated on-chip lasers~\cite{contu20143d}, etc. 
However, silicon's inherently narrow bandgap restricts its applications beyond telecommunications band, especially in fields that require shorter wavelengths, such as atomic physics, augmented and virtual reality, and bio-sensing~\cite{lu2024emerging}. Heterogeneous integration with wideband materials is a straightforward way to bypass this limitation~\cite{guo2023ultra}.
So far, 3D PICs on silicon substrates in ultraviolet and visible bands have been developed through vertical integration with silicon nitride (SiN)~\cite{sorace2019versatile,lin2021low}, aluminum oxide (Al$_2$O$_3$)~\cite{west2019low,mu2019monolithic,mu2019resonant}, and sputtered aluminum nitride (AlN)~\cite{dong2022piezo,zhu2015vertically}. These have been implemented for ion trapping~\cite{sorace2018multi}, light detection~\cite{lin2022monolithically}, piezo-actuated modulators~\cite{dong2022high}, etc.

\begin{figure*}[htbp]
\centering
\includegraphics[width = \textwidth]{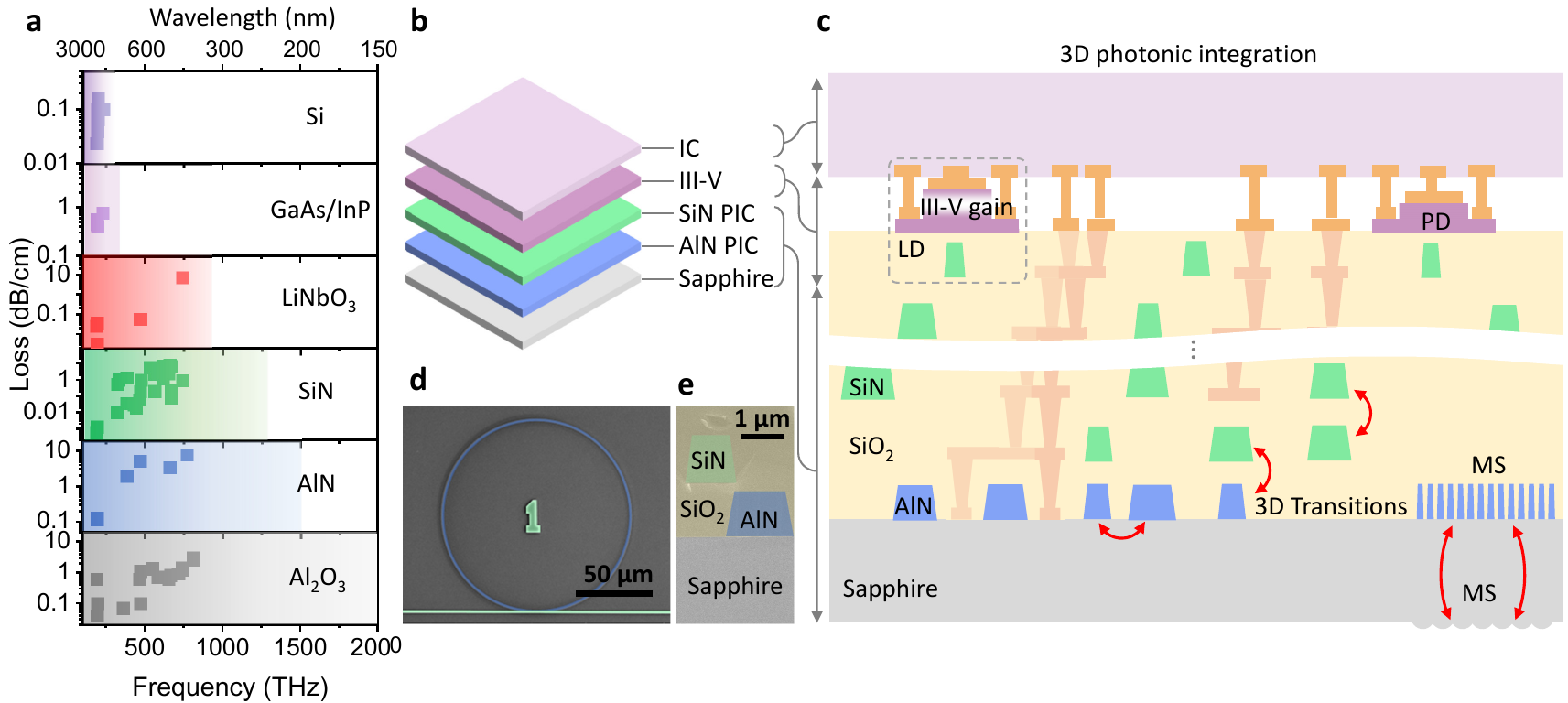}
\caption{\textbf{3D heterogeneous PICs}. \textbf{a}, Comparison of optical loss and transparent window (color-shading area) between popular integrated platforms and the materials used in this work. 
\textbf{b}, Concept of 3D heterogeneous PICs.
\textbf{c}, Cross-section of this 3D PICs in \textbf{b}. That is envisioned future works with extended functionality, such as control ICs, III-V laser diodes (LDs) and photon detectors (PDs), scale-up multilayer heterogeneous SiN-AlN PICs and other optical microstructures. 
\textbf{d}, False-coloured scanning electron microscope (SEM) image of vertical integrated  AlN microring and SiN waveguide.
\textbf{e}, False-coloured SEM image of cross-section for stacked photonics waveguides on sapphire substrate.}
\label{fig:figure1}
\end{figure*}

Compared to popular integrated materials such as silicon (Si), gallium arsenide/indium phosphide (GaAs/InP) and lithium niobate (LiNbO$_3$), SiN, Al$_2$O$_3$ and AlN have much broader optical windows spanning from ultraviolet to infrared wavelengths. Notably, they are transparent in the ultraviolet spectrum (wavelengths below 400 nm), a region inaccessible to most widely used integrated materials. This property is highly valued for current short-wavelength integrated photonics~\cite{ludwig2024ultraviolet}. Furthermore, they exhibit low optical propagation losses, comparable to those of other materials within corresponding spectral bands. For clarity, we summarized the reported optical propagation losses in waveguides of these materials in their transparent spectral windows in Figure\;\ref{fig:figure1}a~\cite{biberman2012ultralow,tran2018ultra,wilmart2020complete,ciminelli2013high,chang2020ultra,chang2019low,apiratikul2014enhanced,sugimoto2004low,li2023high,li2023high,zhang2017monolithic,gao2022lithium,hwang2023tunable,puckett2021422,yong2022power,subramanian2013low,smith2023sin,chauhan2020ultra,corato2024absorption,morin2021cmos,buzaverov2023low,ji2017ultra,chanana2022ultra,liu2018ultra,lu2018aluminum,liu2021aluminum,bruch201817,sun2019ultrahigh,liu2022fundamental,sorace2018multi,west2019low,mckay2023high,demirtacs2018low,de20192,bradley2007fabrication,aslan2010low}. The optical losses tend to increase at higher frequencies due to stronger scattering caused by waveguide surface roughness and intrinsic material absorption. Technological advancements in material preparation and nanofabrication processes hold promise for substantially minimizing these losses, potentially approaching the low loss levels observed in the infrared band. Due to broadband and low-loss characteristics, SiN, AlN, and Al$_2$O$_3$ are excellent candidates for ultra-wideband 3D heterogeneous PIC systems.

In this work, we developed a 3D PIC platform by vertically integrating SiN and AlN PICs on sapphire substrate, offering ultra-broad optical transparency from ultraviolet to infrared spectrum. To evaluate the platform's robustness, we demonstrated both nonlinear and linear optical processes in visible (VI) and near-infrared (IR) spectral regions with fundamental components of the platform. A low-loss AlN microring, vertically coupled with a SiN waveguide, was fabricated, achieving intrinsic quality factors of 2.36$\times$10$^6$ and 3.35$\times$10$^5$ for near-IR and VI wavelength bands, respectively. This enabled efficient second harmonic generation (SHG) with an efficiency of 12598 $\%$/W and spontaneous parametric down-conversion (SPDC) with a single-photon generating rate of 24 MHz/mW. We also built essential SiN optical components of the platform, operating in both IR and VI ranges, including high-Q microcavities, low-loss waveguide spirals, wavelength division multiplexers, and tunable Mach-Zehnder interferometers. Additionally, vertical coupling between AlN and SiN PICs was also demonstrated. Our work presents a 3D SiN-AlN PICs with high-performance linearity and nonlinearity, showing great potential for integrated quantum photonics. 

\section{Results}

\subsection{Ultra-wideband 3D PICs}

The envisioned 3D PICs feature a layered architecture as shown in Figure\;\ref{fig:figure1}b, consisting of ICs, a III-V gain layer, SiN and AlN PIC layers, and a sapphire substrate with integrated microstructures (MS). A typical cross-section is illustrated in Figure\;\ref{fig:figure1}c. 
ICs serve as driving and control units, connected to PICs via wafer bonding~\cite{reif2002fabrication,ranno2023multi,chang20233d}. III-V materials provide active optical components like laser gain media and photodetectors~\cite{franken2023hybrid,op2021iii}. 
SiN waveguides have low optical losses spanning a broad spectrum from infrared to ultraviolet. Meanwhile, their vertical stacking is highly compatible with standard COMOS fabrication processes, positioning them as an ideal platform for photon routing in 3D PICs. However, the centrosymmetric structure of SiN, inherently limits the critical functionalities of 3D PICs, such as efficient second-order optical nonlinearity and high-speed electro-optic modulation. AlN, featuring a high index (2.2) and large bandgap (6.2 eV), supports tightly confined, low-loss waveguides over ultra-wide band~\cite{li2021aluminium}. Moreover, AlN possesses strong $\chi^{(2)}$ nonlinearity, electro-optical effect, and intrinsic piezoelectricity, enabling essential photonic operations within the 3D PICs, including wavelength conversion via $\chi^{(2)}$ non-linearity and high-speed optical modulations~\cite{li2021aluminium}.
The sapphire substrate, with a large bandgap (7.6 eV), provides an ultra-wide transparent window and is widely used in optical microstructures~\cite{dobrovinskaya2009sapphire}. 
Interlayer optical transitions can be achieved via evanescent coupling between adjacent layers or through microstructures. Optical modulation and control circuits are linked through interlayer metals, similar to interconnections in ICs.

Foundation of our ultra-wideband 3D PIC concept is the heterogeneous integration of SiN and AlN. This approach exploits the fabrication compatibility of SiN and  nonlinearity of AlN, thereby supporting the development of high-density, large-scale 3D PICs with advanced photonic functionalities. In this research, we focus on the integration of AlN and SiN PICs on sapphire substrate, using SiO$_2$ as cladding layer. The detailed fabrication process was described in the following section. A typical 3D PIC structure as shown in Figure\;\ref{fig:figure1}d (captured during fabrication process), illustrates an AlN micro-ring (blue) on sapphire, surrounded by SiO$_2$, with a SiN bus waveguide (green) positioned above it. The cross-section of vertically integrated SiN and AlN waveguides is depicted in Figure\;\ref{fig:figure1}e.

\begin{figure*}[htbp]
\centering
\includegraphics[width=\textwidth]{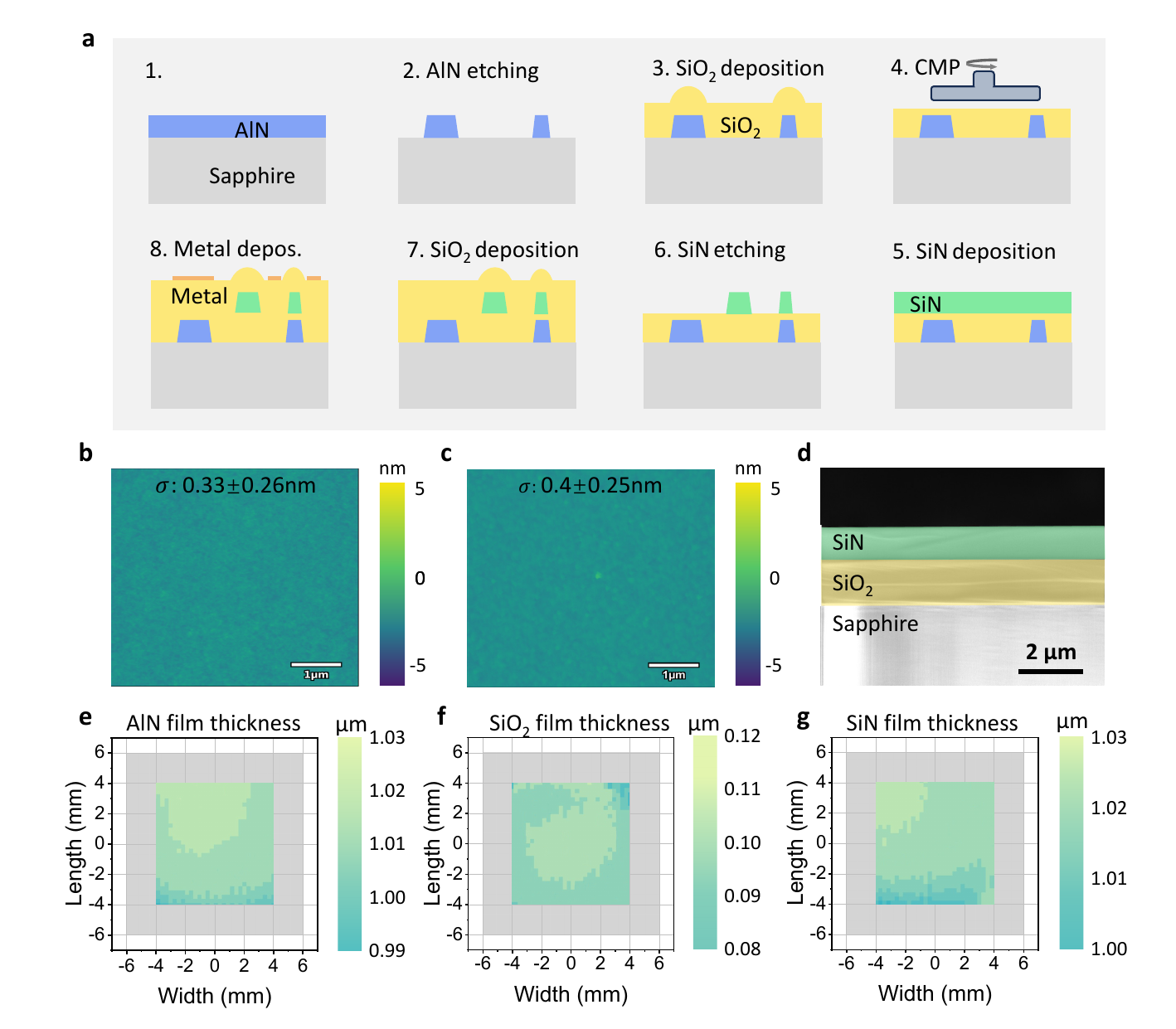}
\caption{ \textbf{Fabrication of 3D PICs}. \textbf{a}, Fabrication process flow of vertical heterogeneous integration of AlN and SiN PICs. \textbf{b}, Two dimensional (2D) atomic force microscopy (AFM) picture of polished SiO$_2$ surface with the root-mean-squared roughness ($\sigma$) of 0.33 $\pm$ 0.26 nm. \textbf{c}, 2D AFM picture of SiN surface with a $\sigma$ of 0.4 $\pm$ 0.25 nm. \textbf{d}, False-coloured scanning electron microscope (SEM) image of cross-section of integrated multilayer on sapphire. Thickness mapping of AlN thin film (\textbf{e}), SiO$_2$ thin film deducting AlN thickness (\textbf{f}), and SiN thin film (\textbf{g}).}
\label{fig:figure2}
\end{figure*}

\subsection{Fabrication of 3D PICs}

The 3D PICs were fabricated following the process flow outlined in Figure\;\ref{fig:figure2}a. Initially, the AlN waveguide structures were patterned using electron beam lithography (EBL) with Fox-16 resist. Following development in the TMAH developer, ICP dry etching with Cl$_2$/BCl$_3$/Ar gases was used to etch AlN, transferring the patterns into the AlN layer.
And then, a 1.5-$\mu$m-thick SiO$_2$ was deposited by PECVD to cover AlN waveguides. A 3-hour annealing at 1000$^{\circ}$C in nitrogen atmosphere was performed to minimize the optical absorption of SiO$_2$ cladding layer. 
The SiO$_2$ surface was planarized by chemical mechanical polishing (CMP) using Fujmi PL-4217 Slurry and IC-1000 Polish pad. The polished SiO$_2$ surface has a roughness ($\sigma$) of 0.33 $\pm$ 0.26 nm (Figure\;\ref{fig:figure2}b).
Subsequently, a 1-$\mu$m-thickness SiN layer was deposited on the flattened SiO$_2$ by LPCVD (Figure\;\ref{fig:figure2}d), yielding a surface roughness $\sigma$ of 0.4 $\pm$ 0.25 nm (Figure\;\ref{fig:figure2}c). 
We used MaN2410 resist to define SiN PICs patterns by EBL. After developing in MIF319 developer, HCF$_3$/O$_2$ chemistry was used to etch SiN.
Afterwards, SiN PICs were annealed at 1000$^{\circ}$C in nitrogen to reduce the optical absorption of SiN waveguides. Another 1.5-$\mu$m-thickness SiO$_2$ cladding layer was deposited and annealed following previous processes. Finally, Ti/Au metal stacked layers were deposited as the phase shifter to control the SiN and AlN PICs. We monitored the thickness of AlN, SiO$_2$ and SiN films during fabrication process. As depicted in Figure\;\ref{fig:figure2}e, f and g, these films exhibited thickness variations within approximately $\pm$ 10 nm, indicating a commendable level of uniformity.

\begin{figure*}[htbp]
\centering
\includegraphics[width = \textwidth]{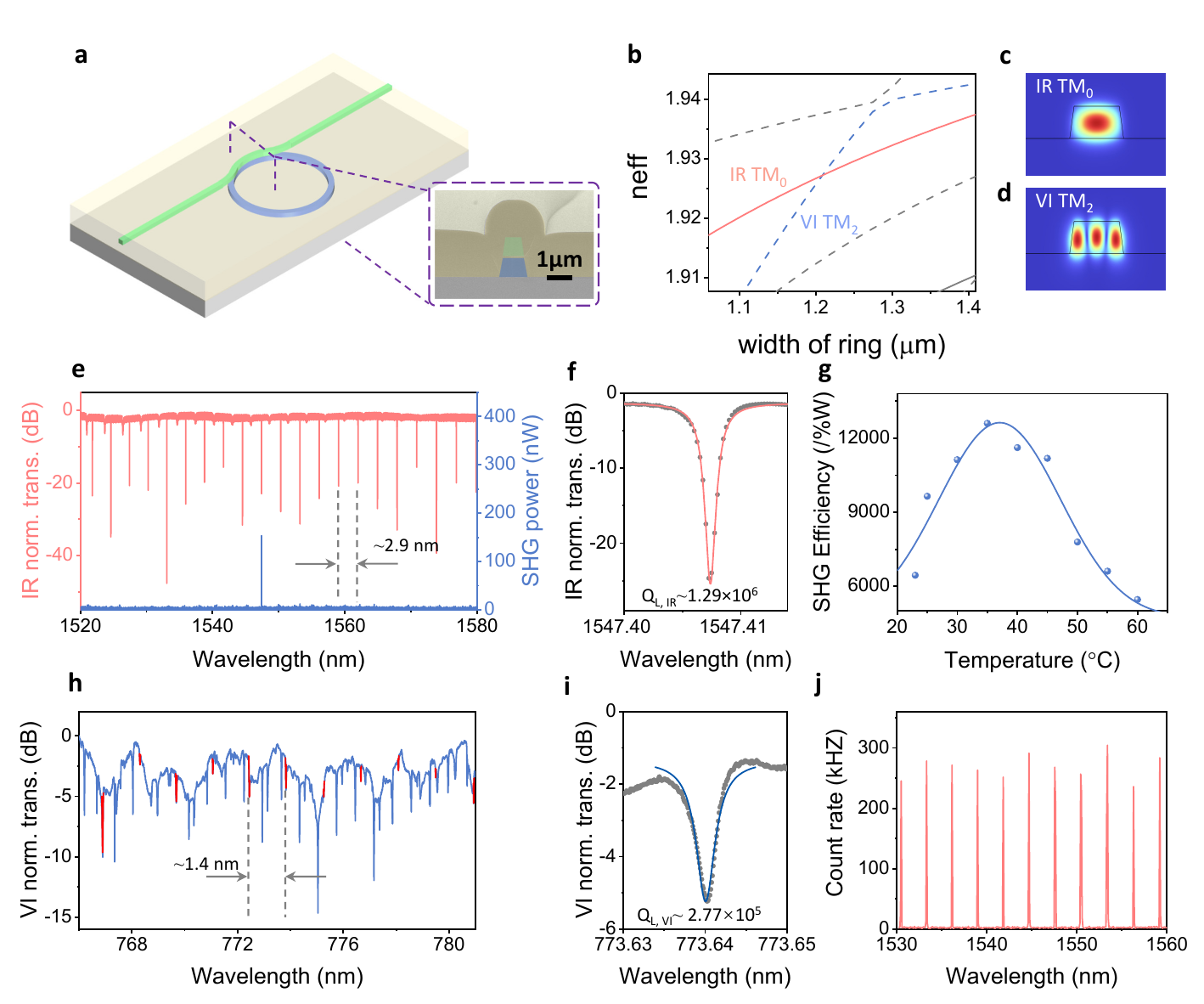}
\caption{ \textbf{SHG and SPDC of AlN PICs}. \textbf{a}, Heterogeneously integrated AlN micro-ring with a pully SiN waveguide. Inset shows a false-colored SEM image of cross-section of coupling area.
\textbf{b}, Simulated optical effective refractive index (n$_{eff}$) of AlN ring waveguide for infrared (IR, red) and visible (VI, blue) modes.
\textbf{c} and \textbf{d}, Optical IR TM$_0$ and VI TM$_2$ in AlN ring waveguide.
\textbf{e}, IR transmission spectrum (red) and generated second harmonic generation (SHG) spectrum (blue) of AlN ring.
\textbf{f}, Zoom-in view of the IR resonance in \textbf{e}, aligned with the SHG peak. The loaded quality factors (Q$_{L,IR}$) of this resonance is estimated to be about 1.29$\times$10$^6$ by Lorentzian fitting.  
\textbf{g}, SHG efficiency tuned by temperature. 
\textbf{h}, VI transmission spectrum of AlN ring. TM$_2$ mode resonances are highlighted by red color.
\textbf{i}, Zoom-in view of VI resonance in \textbf{h} aligned with the SHG peak. Its loaded quality (Q$_{L,VI}$) is about 2.77$\times$10$^5$.
\textbf{j}, Single-photon spectrum generated via SPDC of AlN ring.
}
\label{fig:figure3}
\end{figure*}

\subsection{Optical non-linearity of AlN PICs}

On-chip optical nonlinearity plays a crucial role in both classical and quantum photonics information processing. Single-crystalline AlN stands out as a prominent material for nonlinear phenomena across telecom IR, visible, and ultraviolet wavelengths~\cite{zhang2023chip,bruch201817,bruch2021pockels,chen2021supercontinuum,Sliu2019beyond}. To exhibit the nonlinearity of our 3D PICs, we fabricated an AlN micro-ring cavity and demonstrated efficient SHG and SPDC in visible and near-infrared spectra. 
As shown in Figure\;\ref{fig:figure3}a, the micro-ring device comprises a 60-$\mu$m-radius ($R_r$) AlN ring on sapphire with a pulley SiN waveguide thereon. Both AlN ring and SiN layers are about 1 $\mu$m thick, separated by a 100-nm gap and encased in a thick SiO$_2$ cladding layer (inset of Figure\;\ref{fig:figure3}a). Coupling between ring and waveguide was optimized using finite element simulations (Supplementary Section I).      

Efficient SHG and SPDC require phase-matching that can be fulfilled by the fundamental transverse-magnetic IR mode (TM$_0$) and higher-order VI mode (TM$_2$) (Figure\;\ref{fig:figure3}c and d). Their angular frequencies ($\omega_{vi, ir}$) and momentum ($m_{vi, ir}$) relate as $\omega_{vi}=\omega_{ir,1}+\omega_{ir,2}$ and $m_{vi} = m_{ir,1}+m_{ir,2}$, respectively, which require effective refractive indices matching for two modes, $n_{eff} = m_{vi, ir}c/\omega_{vi, ir}R_r$ (where $c$ is the vacuum light speed). Numerical simulations of mode dispersion in AlN ring waveguides were performed to obtain $n_{eff}$. By engineering AlN microring waveguide width, 1550-nm TM$_0$ and 775-nm TM$_2$ reach effective refractive-index matching at a width of $\sim$1.2 $\mu$m (Figure\;\ref{fig:figure3}b). 

Linear optical properties of AlN microring were first characterized via transmission spectra (Supplementary Section IV). The IR spectrum (red) shows a free spectral range (FSR) of 2.9 nm for TM$_0$ mode (Figure\;\ref{fig:figure3}e). A phase-matched IR resonance at 1547.66 nm aligns with SHG peak in VI spectrum (blue), indicating near-critical coupling. Its zoomed-in view is further presented in Figure\;\ref{fig:figure3}f. Its loaded quality factor (Q$_{L,IR}$) and intrinsic quality factor (Q$_{in,IR}$) are 1.29$\times$10$^{6}$ and 2.36$\times$10$^{6}$, respectively, corresponding to an IR optical propagation loss of 0.165 dB/cm. Figure\;\ref{fig:figure3}h shows the visible transmission spectrum of AlN microring, featuring multiple mode families with FSRs around 1.45 nm. These resonances mainly include TM$_0$, TE$_0$ and TM$_2$ modes, where TM$_2$ mode resonances are highlighted in red. The TM$_2$ mode aligning SHG peak of 773.82 nm is displayed in Figure\;\ref{fig:figure3}i, indicating the loaded quality factor (Q$_{L,VI}$) of 2.77$\times$10$^{5}$ and intrinsic quality (Q$_{in,VI}$) of 3.35$\times$10$^{5}$, corresponding to a VI optical propagation loss of 1.2 dB/cm. 

As shown in Figure\;\ref{fig:figure3}a, a 200-nW SHG power was observed with a 0.63-mW input pump power at wavelength of 1547.41 nm. Considering 20 $\%$ and 10 $\%$ edge coupling efficiencies of SiN waveguides for IR and VI modes, respectively, the on-chip SHG efficiency is estimated to be about 6450 $\%$/W. To maximize this efficiency, we precisely aligned IR and VI resonances via thermal-optical effect for perfect phase matching (Supplementary section IV). At 38 $^\circ$C, the SHG efficiency reached maximum value of 12598 $\%$/W (Figure\;\ref{fig:figure3}g), roughly consistent with the theoretical value of 14878 $\%$/W (Supplementary Section II).

Efficient SHG confirms phase-matching of optical modes in AlN microring and also indicates its potential for efficient SPDC. Next, we further demonstrated the single-photon generation by SPDC (Supplementary section IV). A 5-mW pump power launched at the visible TM$_2$ mode resonance at $\sim$ 773.8 nm. Generated single photons were detected by a superconducting single-photon detector (SSPD). Figure\;\ref{fig:figure3}j is the measured single photon spectrum with 13 peaks, which align with 13 IR resonances of AlN microrings, spanning a bandwidth over 30 nm. Considering pump/single photon input/output coupling efficiencies and photon loss ($\sim$6 dB) of measurement system, the single photon generation efficiency is about 24 MHz/mW, closely matching theoretical result of 26 MHz/mW (Supplementary section III).

\begin{figure*}[htbp]
\centering
\includegraphics[width=\textwidth]{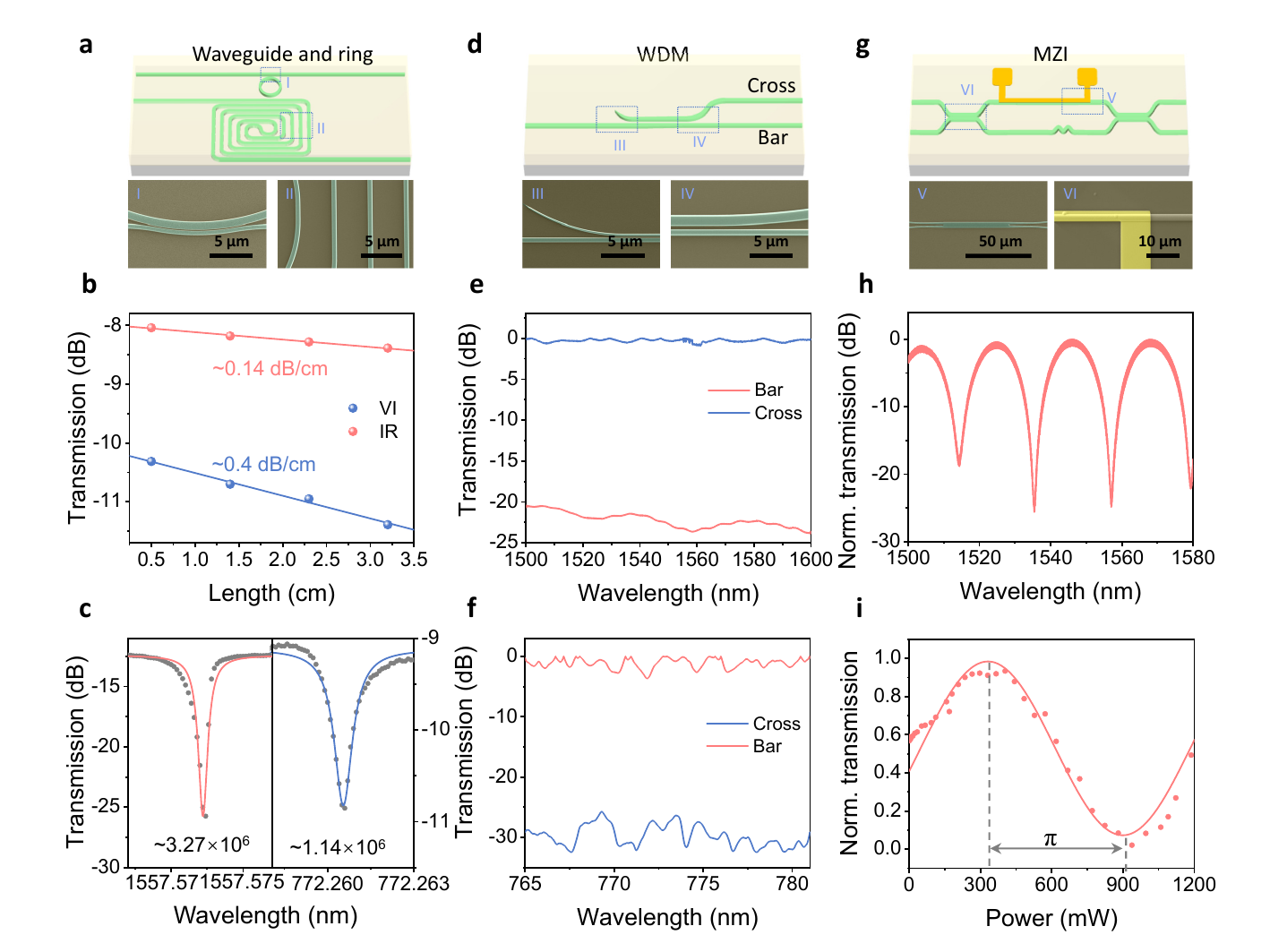}
\caption{\textbf{Optical components of SiN PICs}.
\textbf{a}, Schematic of SiN microring and spiral waveguide on SiO$_2$/sapphire. False-color SEM images of ring (I) and waveguide (II).
\textbf{b}, Measured transmission intensities of SiN spiral waveguides with different lengths for IR and VI optical bands, respectively. 
\textbf{c}, Optical SiN micro-ring resonances with intrinsic quanlity factors of 3.27$\times$10$^6$ and 1.14$\times$10$^6$ for IR and VI TM$_0$ modes, respectively.
\textbf{d}, Schematic of WDM and its zoom-in area (False-color SEM images (III) and (IV).
\textbf{e} and \textbf{f}, WDM transmission spectra of bar (blue) and cross (red) ports in IR and VI optical bands.
\textbf{g}, Schematic of Mach–Zehnder interferometer (MZI) with thermal phase shifter on its one arm. False-color SEM images of MMI (V) and electrode (VI).
\textbf{h}, Normalized transmission spectrum of MZI. 
\textbf{i}, Normalized transmission intensities of MZI tuned by thermal power with input wavelength at 1550 nm.  
 }
\label{fig:figure4}
\end{figure*}

\subsection{Optical components of SiN PICs}

In our 3D PICs, SiN waveguides hold a key position for photon guidance and manipulation, which typically require low optical propagation loss for SiN waveguide-based optical components. To assess these losses, we fabricated SiN spiral waveguides (1 $\mu m$ $\times$ 1.2 $\mu m$ cross-section) on SiO$_2$/sapphire, as shown in Figure\;\ref{fig:figure4}a. Transmission intensities of TM$_0$ modes of waveguides with different lengths are plotted as dots in Figure\;\ref{fig:figure4}b. By linearly fitting these data points, we extracted propagation losses of 0.14 dB/cm and 0.4 dB/cm for IR and VI wavelength bands, respectively. Additionally, an SiN microring was also fabricated as presented in Figure\;\ref{fig:figure4}a, with a 60-$\mu$m radius and a cross-section of 1 $\mu m$ $\times$ 1.2 $\mu m$. This microring exhibited the typical intrinsic quality factors of 3.27$\times$10$^{6}$ (at $\sim $1550 nm) and 1.14$\times$10$^{6}$ (at $\sim $775 nm) for TM$_0$ modes (Figure\;\ref{fig:figure4}c), corresponding to losses of 0.12 dB/cm and 0.33 dB/cm, which are consistent with the estimates of spiral waveguides. 

Wave division multiplexing (WDM) is a key component in integrated optics. In our 3D PICs, the SiN WDM functions to combine and separate VI and IR light branches. As shown in Figure\;\ref{fig:figure4}d, the SiN WDM comprises two closed parallel taper waveguides with a 600-nm coupling gap and 1-mm coupling length. The device is optimized for optical TM$_0$ modes in both VI and IR bands, and the design details can be found in Supplementary Section V. Figure\;\ref{fig:figure4}e is the calibrated WDM transmission spectrum in IR band. The IR light almost entirely couples into the cross port, while only smaller part ($<$ -20 dB) passes through the bar port. Meanwhile, the insertion loss is only about 0.3 dB. Conversely, for the VI band,  the bar port has near unitary transmission with minimal transmission of the cross port ($<$ -25 dB), and the corresponding insertion loss is 1.1 dB. 

Mach-Zehnder Interferometer (MZI) is a typical component used for controlling photon routing in PICs. To demonstrate this functionality in our 3D PICs, a SiN unbalanced MZI was fabricated with a specific design as depicted in Figure\;\ref{fig:figure4}g. This MZI comprises two 50/50 multimode interference (MMI) couplers, and a phase shifter is placed on one arm. Due to symmetrical material structure, SiN hasn’t an electro-optic coefficient. So, we use a heater as the phase shifter to tune the phase of optical modes via thermal-optic effect. SiN has a thermos-optic coefficient of 4.7$\times$10$^{-5}$/K, smaller than that of Si (1.84$\times$10$^{-4}$/K).
Numerical simulations were conducted to optimize the MMI structure parameters (Supplementary Section V). The IR transmission spectrum of unbalanced MZI is shown as Figure\;\ref{fig:figure4}h, exhibiting a typical sinusoidal oscillation with an extinction ratio over 25 dB. Furthermore, we fixed the input optical wavelength at 1550 nm and controlled light output route and output power by fine-tuning MZI phase via the heater. Figure\;\ref{fig:figure4}i shows the normalized output power at one MZI port, depending on heater's driving power. A $\pi$ phase shift was achieved with about 600-mW power. 

\subsection{Optical link across AlN and SiN layers}

\begin{figure}[htbp]
\centering
\includegraphics[width=0.5\linewidth]{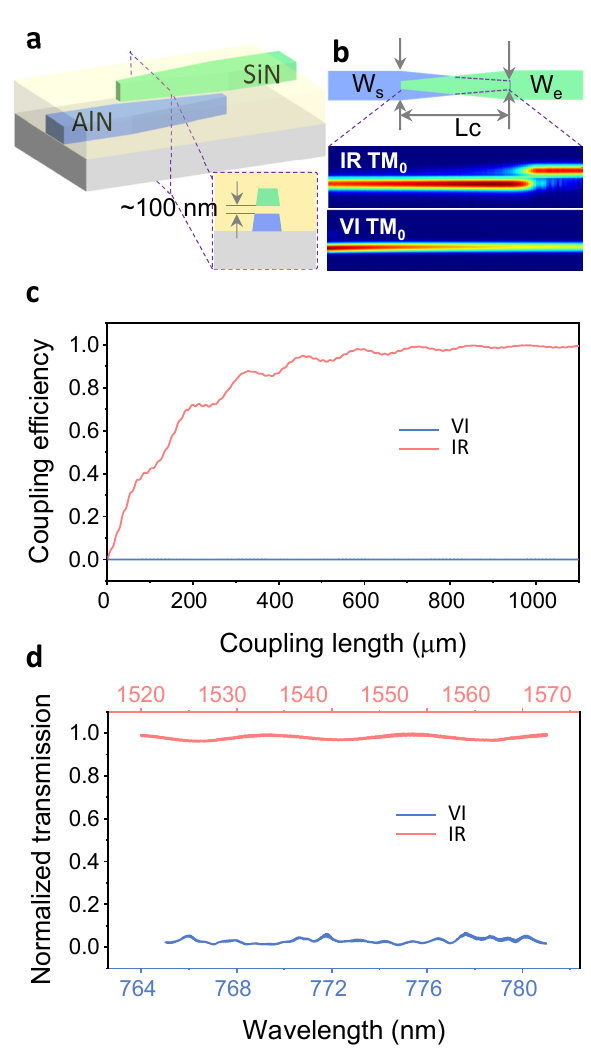}
\caption{\textbf{Optical link between AlN and SiN PICs.}
\textbf{a}, Schematic vertical coupler consists of AlN and SiN taper waveguides separated by a gap in vertical direction. \textbf{b}, Simulated mode profile distribution along directional coupler for IR and VI TM$_0$ modes. \textbf{c}, Simulated coupling efficiency from AlN to SiN as a function of the coupler length ($L_c$). \textbf{d}, Measured transmission spectrum of the coupler.
}
\label{fig:figure5}
\end{figure}

Besides high-performance devices, reliable optical interconnects between stacked photonic layers are crucial for 3D PICs. Here, we proposed an adiabatic taper coupler to link SiN and AlN PIC layers as illustrated in Figure\;\ref{fig:figure5}a. The coupler was designed with a 100-nm coupling gap, 100-nm tip width ($W_e$), 1.2-$\mu$m waveguide width ($W_s$) and a coupling length of $L_c$ (Figure\;\ref{fig:figure5}b). Numerical simulation was conducted to optimize $L_c$ for maximizing the coupling efficiency of coupler. IR TM$_0$ mode of 1550 nm and VI TM$_0$ mode of 775 nm were selected as the operating optical modes based on our previous components of 3D PICs. The dependence of simulated coupling efficiency on coupling length $L_c$ is illustrated as Figure\;\ref{fig:figure5}c. For IR mode, the coupling efficiency increases with $L_c$ and saturates near-unity at $L_c$ $>$ 1000 $\mu$m. While the VI mode has weak coupling between AlN and SiN taper waveguides, forbidding efficient optical crossing. Mode profile distributions within coupler waveguides for two optical modes are shown as in Figure\;\ref{fig:figure5}b. Calibrated optical transmission spectra of coupler device for IR and VI wavelength bands are plotted in Figure\;\ref{fig:figure5}d, presenting near-unity and near-zero coupling efficiencies for IR and VI modes, respectively. To achieve efficient vertical coupling for shorter wavelength bands, further optimizing coupler structures is necessary. Additionally, alternative methods such as vertical MMIs~\cite{brooks2011vertically}, intermediate directional couplers~\cite{takei2015low}, and grating couplers~\cite{sodagar2014high}, offer viable solutions for realizing efficient broadband coupling.

\section{Discussion}

AlN, SiN, and Al$_2$O$_3$ have ultra-wideband optical transparency from ultraviolet to infrared, far beyond that of silicon~\cite{sorace2018multi,Sliu2019beyond,lu2018aluminum,aslan2010low,li2021aluminium}. This positions them as ideal candidates for emerging applications of PICs working at short wavelengths~\cite{lu2024emerging}. Leveraging these three materials, we have developed 3D PICs via heterogeneous integration and demonstrated efficient SHG and SPDC, and low-loss and tunable optical components. Moreover, optical linking across vertical stacking layers was explored. We also assessed the ultra-low crossing loss between adjacent phonics layers by numerical simulation (Supplementary Section VI). Although the 3D PICs demonstrated promising performance of both optical non-linearity and linearity, there still remains significant room for improvement. First, the device fabrication processes need to be optimized to further minimize optical losses. For example, SiN PICs have already demonstrated optical losses as low as 1 dB/m and 0.1 dB/m at VI and IR wavelength bands~\cite{puckett2021422,chanana2022ultra,ji2017ultra,yong2022power}, respectively, much better than our current results. Second, the structures of 3D PICs should be further optimized to achieve broadband optical vertical coupling. This may involve replacing the SiO$_2$ cladding material with Al$_2$O$_3$, reducing the thickness of AlN and SiN waveguide, and exploring other coupling structures such as vertical multimode interferometers and grating couplers~\cite{lin2021low,mu2019resonant,lin2022monolithically,shang2015low}. Finally, other optical components working at ultraviolet to visible bands remain to be unexplored. 



\section{Conclusion}

In conclusion, we have developed a 3D PIC platform based on the vertical heterogeneous integration of AlN and SiN layers, enabling broadband operation across wavelengths from ultraviolet to infrared. The platform’s efficient optical nonlinearity was validated through SHG and SPDC experiments performed in an AlN microring. Furthermore, we fabricated high-performance SiN optical components, such as low-loss microrings and waveguides, WDM devices, and tunable MZIs, demonstrating their immense potential for applications in 3D PICs. Optical interlayer transitions between AlN and SiN PICs were achieved via evanescent-field coupling between closely spaced waveguides. The scalability and integration density of our 3D PICs can be further enhanced by vertically stacking additional SiN waveguide layers. Additionally, thin III-nitride films, capable of photon generation and detection across ultraviolet to visible wavelengths~\cite{razeghi2011iii}, can be directly integrated onto the PICs, thereby enabling the incorporation of numerous off-chip optical devices into a single chip. The integration of III-nitride-based ICs also offers a pathway to co-developing a 3D electronic-photonic integrated circuit ecosystem~\cite{xiang2024building,zhang2020scalable}, potentially unlocking new possibilities for both classical and quantum information processing.

\section{Methods}

Device fabrication process can be found in the main text. Device measurements are detailedly described in supplementary materials. 
\begin{acknowledgement}
LZ, YG, JY, JW and JL acknowledge the support from National Natural Science Foundation of China (62135013, 62274163, 62250071, 62234001); Beijing Nova Program 20230484466; Youth Innovation Promotion Association CAS 2022000028 and 2023123.
\end{acknowledgement}

\section{Author contributions}

The experiments were conceived by LZ. The device was designed and fabricated by LZ. Measurements were performed by LZ. Analysis of the results was conducted by LZ. YG, JY, JW and JL support this work. All authors contribute to the preparation of manuscript. 

\section{Competing interests}

The Authors declare no Competing Financial or Non-Financial Interests

\section{Data availability}

The data that support the findings of this study are available from the corresponding author upon reasonable request

\begin{suppinfo}

Vertical coupling design of AlN microring, theory of SHG and SPDC in microcavity, measurement of SHG and SPDC, design of SiN WDM and MMI, loss evaluation of vertical waveguide crossings

\end{suppinfo}

\bibliography{achemso-demo}

\end{document}